\documentclass[twocolumn,showpacs,aps,prl,
groupedaddress,amssymb,amsmath,nobalancelastpage]{revtex4}

\usepackage{graphicx}
\usepackage{longtable}

\begin{document}

\title{Limits of the equivalence of time and ensemble averages in shear flows.}

\author{Yuhong Wang}
\author{Kapilanjan Krishan}
\author{Michael Dennin}
\affiliation{Department of Physics and Astronomy, University of
California at Irvine, Irvine, California 92697-4575}

\date{\today}

\begin{abstract}
In equilibrium systems, time and ensemble averages of physical
quantities are equivalent due to ergodic exploration of phase
space. In driven systems, it is unknown if a similar equivalence
of time and ensemble averages exists. We explore effective limits
of such convergence in a sheared bubble raft using averages of the
bubble velocities. In independent experiments, averaging over time
leads to well converged velocity profiles. However, the
time-averages from independent experiments result in distinct
velocity averages. Ensemble averages are approximated by randomly
selecting bubble velocities from independent experiments.
Increasingly better approximations of ensemble averages converge
toward a unique velocity profile. Therefore, the experiments
establish that in practical realizations of non-equilibrium
systems, temporal averaging and ensemble averaging can yield
convergent (stationary) but distinct distributions.

\end{abstract}

\maketitle

A central tenant of statistical mechanics is the ergodic
hypothesis: during time evolution, thermodynamic systems pass
through almost all possible microstates leading to the equivalence
of time and ensemble averages
\cite{Boltzman_1871c,Boltzman_1895a,Boltzman_1895b,LandauLifshitz,
Pathria}. The exploration of phase space is achieved through
thermal fluctuations, and this allows for a description of
equilibrium dynamics in terms of well defined averaged
observables. The observables are extracted either by averaging
over states explored over long periods of time or from ensembles
of states distributed over all of phase space, whichever method is
most convenient for the observable being studied. A natural
questions is whether or not athermal fluctuations in a driven
system lead to the equivalence of time and ensemble averages
through a similar exploration of phase space. Ideally, testing the
equivalence of time and ensemble averages in driven systems
requires measuring the instantaneous value of an observable for an
infinite period of time (temporal averages) and constructing an
infinite number of ensembles (ensemble averages). This is
untenable for real experiments; however, if time scales of
observation are faster than intrinsic timescales, it is possible
to gather a reasonable approximation of the distribution of values
for an observable.

The averaging process is governed by the fluctuations that move
the system through phase space. In driven, complex fluids,
athermal fluctuations arise through particle rearrangements that
are induced by flow. The potential for these athermal fluctuations
to play a role that is similar to thermal fluctuations in normal
fluids is the basis of many statistical treatments of highly
non-equilibrium complex fluids \cite{CKP97,OODLLN02,BB02,ZBCK05}.
A specific example of this are the various proposals for using
effective temperatures \cite{CKP97,OODLLN02,BB02} in driven
systems. An open question in this field is the quantitative
disagreement \cite{OLN04} between different definitions of
effective temperatures \cite{ZBCK05}. These disagreements may be
connected to the ability of athermal fluctuations to provide
sufficient exploration of phase space for certain variables, such
as velocity. In this letter, we specifically focus on the ability
of athermal fluctuations in the system to explore phase space in a
manner that results in an equivalence between time and ensemble
averages. Testing such an equivalence is an important step in
understanding how well the various elements of statistical
mechanics (such as temperature) can be translated to driven
systems.

In considering any comparison of time and ensemble averages, it is
important to recognize the separate question of the {\it
convergence} of either time or ensemble averages. Independent of
whether the two averages agree, it is possible that the nature of
athermal fluctuations in a driven system are such that average
quantities are not well-defined on experimentally accessible time
scales or number of ensembles. This is highlighted by work in
granular matter in which extremely long times were required for
convergence of the average density under tapping \cite{NKBJN98}.
As we will demonstrate, for our system, both the time and ensemble
averages are well-defined.

In this Letter, we report on measurements of the average velocity
profile using a model two-dimensional foam: a bubble raft
\cite{BL49}. One advantage of this system is the irrelevance of
thermal fluctuations due to the macroscopic nature of the bubbles.
Instead, the fluctuations are the result of bubble rearrangements
during flow \cite{WH99}. Even though the average velocity profile
for foam has been the subject of substantial theoretical
\cite{VBBB03,XOK05a,KD03,WJH06} and experimental
\cite{CRBMGH02,LCD04,DTM01,GD99,RCVH03,DEQRAG05} work, the
connection between time and ensemble averages for velocity
profiles remains an open question.

The bubble raft is produced by flowing regulated nitrogen gas
through a needle into a homogeneous solution of 80\% by volume
deionized water, 5\% by volume Miracle Bubbles (from Imperial Toy
Corporation), and 15\% by volume glycerine. The glycerine provides
additional stabilization of the bubble raft as a function of time.
The bubbles are confined between two parallel bands separated by a
distance $d=57\ {\rm mm}$. The bands are driven at a constant
velocity $v_w$ in opposing directions. This applies a steady rate
of strain to the system given by $\dot{\gamma} = 2v_w/d = 1.4
\times 10^{-3}\ {\rm s^{-1}}$. The total applied strain is $\gamma
= \dot{\gamma}t$, where $t$ is the time interval under
consideration. The direction of flow imposed by the bands is taken
to be parallel to the x-axis. We use a CCD camera to capture the
state of the system at a frame rate of $10\ {\rm Hz}$. A Particle
Image Velocimetry program \cite{piv} is then used to extract
velocities of individual bubbles using consecutive images. Each
consecutive pair of images corresponds to a measurement of its
instantaneous velocity. For this paper, we focus exclusively on
the behavior of the velocity in the x-direction, and the system
was divided into forty bins in the y-direction so as to measure
the velocity as a function of y. The bubbles are relatively
monodisperse, with a bubble diameter of $D = 2.66 \pm 0.2\ {\rm
mm}$. Details of the apparatus and methods can be found in
Ref.~\cite{wkd05a}.

The results presented in this letter are based on thirty different
experimental realizations of a bubble raft. All of the parameters
for each experimental realization were the same except for the
configuration of the bubbles. In each of the experimental
realizations, the system is subjected to a total strain of five
and $N= 1000$ consecutive images of the bubbles are captured. From
these thirty different experimental realizations, we develop two
different data sets: a time set and an ensemble set. The time set
is the 30 independent time series of the bubble velocities. A
member of the time set corresponds to a single experimental
realization with $1000$ elements. Each element captures the
instantaneous position and velocity of the bubbles as measured
between successive images at a given time (or value of strain). A
member of the ``ensemble set" is constructed from the thirty
experimental realizations by randomly sampling $1000$ elements
from the thirty experimental realizations. We build 30 such
members to constitute the ensemble set. This procedure ensures an
equal number of elements in each of the members of the time and
ensemble sets. This allows for a direct statistical comparison
between the two data sets. A time/ensemble average refers to
averaging all instantaneous velocities from a individual member of
the time/ensemble set of data.

\begin{figure}
\includegraphics[width=8.6cm]{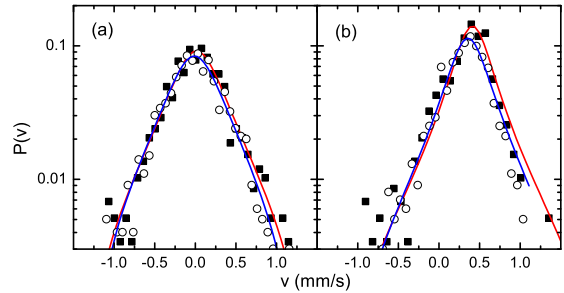}
\caption{(color online) The probability distribution of velocities
is indicated for ensemble averaging (open circles) and time
averaging (closed squares). The solid lines are Lorenztian fits to
the data, red curves for time averaged data and blue for ensemble
averaged data. (a) Results for a position at the center of the
trough. For the ensemble (time) average the center of the
Lorentzian fits are at $0.01 \pm 0.01$ ($-0.01 \pm 0.01$) and the
widths are $0.68 \pm 0.02$ ($0.72 \pm 0.02$), respectively. (b)
For a position $0.25d$ from the center. For the ensemble (time)
average the center of the Lorentzian fits are at $0.41 \pm 0.01$
($0.36 \pm 0.01$) and the widths are $0.47 \pm 0.03$ ($0.52 \pm
0.02$), respectively. } \label{prob_distributions}
\end{figure}

To characterize the system's behavior, we measured the
distribution of velocities, the time-correlation of the
velocities, the average velocity profiles, and characterized the
convergence rate of the averaging process. All of these studies
(except the velocity correlations) were made for both the time and
ensemble data sets. The results for the velocity distributions are
shown in Fig.~\ref{prob_distributions} for a single member of both
the time and ensemble set. Sampling the velocities from either
time sets \cite{wkd06b} or ensemble sets result in a Lorentzian
distribution. However, as we will show, a key distinction lies in
the distribution of the mean values for different members of the
respective sets.

\begin{figure}
\includegraphics[width=8.6cm]{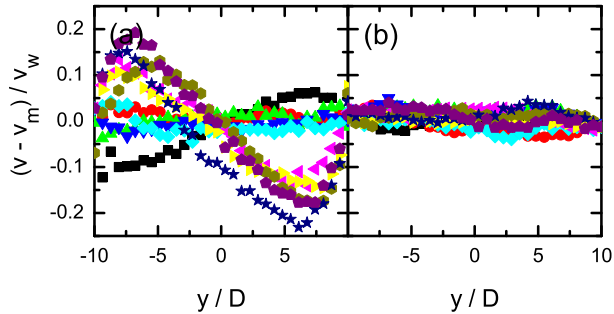}
\caption{(a) Deviation of time average velocity profiles
(normalized by the wall speed $v_w$) from the mean profile ($v_m$)
generated by ten members of the time set. (b) Deviation of ten
different ensemble averages of the velocities from the mean
profile. Plotting the deviation highlights the differences between
the different time average profiles. The average standard
deviation for time average profiles in (a) is $0.081$, in
contrast, the ensemble averages shown in (b) are relatively
indistinguishable with a standard deviation of $0.017$. The
y-position is scaled by the bubble diameter $D$.}
\label{vel_profiles}
\end{figure}

The average velocity profiles are approximately linear
\cite{wkd05a}. To characterize the relative deviations of the
various velocity profiles in different sets,
Fig.~\ref{vel_profiles} plots the difference between the
individual average velocity profiles and the mean profile ($v_m$)
for ten members of the time set and ten members of the ensemble
set. In the case of time averaging, the resulting average velocity
profiles qualitatively differ from each other and exhibit
relatively (compared to the ensemble profiles) large deviations
from the mean. In contrast, the ensemble averages are essentially
indistinguishable from each other (this is quantified by the
standard deviation between profiles, see Fig.~\ref{vel_profiles}).

\begin{figure}
\includegraphics[width=8.6cm]{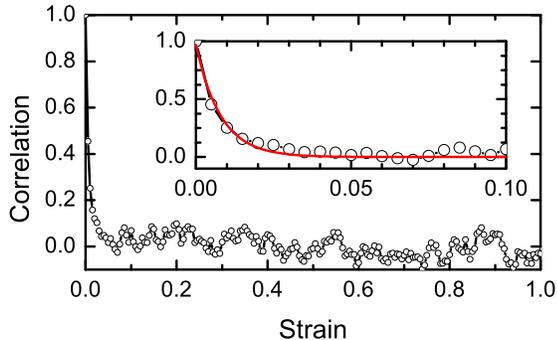}
\caption{This is the decay in the correlation of the mean velocity
$0.25d$ away from the trough center. The total strain in each
experiment is $5$, and we only show the autocorrelation to a unit
strain to make it visually easier to read. The remainder of the
curve is qualitatively unchanged at higher strains. The inset shows
a closeup at short strain along with an exponential fit (correlation
$\propto \exp(-\gamma/0.01)$.}\label{correlation}
\end{figure}

\begin{figure}
\includegraphics[width=8.6cm]{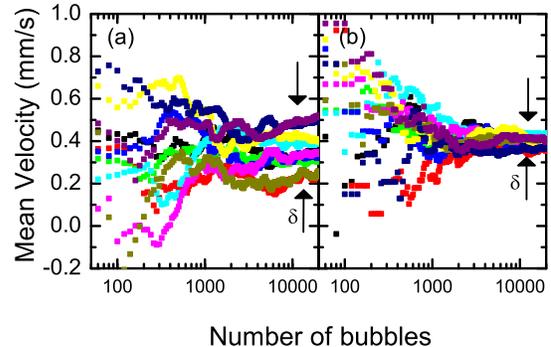}
\caption{(color online) The mean values of the velocity $0.25d$
from the center as a function of the number of bubbles used to
compute the average for (a) time averaging and (b) ensemble
averaging. For illustration purposes, 10 of the 30 runs are shown.
The deviation in the final values (indicated by the arrows and
$\delta$) is larger for time averages than ensemble averages. When
we use the complete set of 30 runs, we obtain $\delta = 0.258{\rm
mm/s}$ for time averages and $\delta = 0.052{\rm mm/s}$ for
ensemble averages.} \label{mean_values}
\end{figure}

There are two key issues regarding the average profiles: (1) are
they computed over independent configurations? and (2) have the
averages converged? Fig.~\ref{correlation} and \ref{mean_values}
address these two questions. In Fig.~\ref{correlation}, the
autocorrelation of instantaneous velocities for a member of a time
set and an exponential fit to the correlation function ($\propto
\exp(-\gamma/0.01)$) are shown. The decay constant corresponding
to a strain of 0.01 is consistent with the previously reported
yield strain in polydisperse bubble rafts ($\sim 0.05$)
\cite{D04}. As the yield strain corresponds to the point at which
elastic response gives way to flow through significant bubble
rearrangements, it is reasonable that this determines the interval
of strain necessary to produce independent configurations.
Therefore, time averages using a total strain of 5 consist of
between 100 to 500 independent configurations. While building an
ensemble member, we randomly choose states of velocity
configurations across all thirty members of the time set. This
results in the average strain between randomly chosen states being
0.15, with a correlation less than the baseline noise in
Fig.~\ref{correlation}. The selection of these states across 30
independent time members, further decreases any correlations
between them.

We find the rate of convergence to the mean value of the velocity
profile is the same for members of the time and ensemble sets.
However, the distribution of final mean values attained are
different, with the time set showing a significantly larger spread
in the converged mean values than those from the ensemble set. In
Fig.~\ref{mean_values}, the convergence of the mean velocity as a
function of the number of bubbles used to compute the average is
plotted for ten members of the time and ensemble averages.
Plotting the mean velocity versus the number of bubbles allows for
direct comparison of the time and ensemble averages. However, it
should be noted, that for the time averages, the number of bubbles
corresponds directly to increasing time, or strain, which in this
case is a strain $\le 5$. In all cases (ensemble and time
averages), the mean values converge to a well-defined value after
averaging over approximately $n = 2000$ bubbles. (This corresponds
to a strain of 1 for the time averages, or approximately 100
independent configurations, and is consistent with previous
measurements of the rate of convergence \cite{LCD04}). The spread
in the final mean values is very different for the two sets. For
the time averages, as expected from Fig.~\ref{vel_profiles}, the
deviation in final mean values is larger than the fluctuation in
the mean for large enough values of $n$. For the ensemble
averages, the deviation in final mean values is similar to the
fluctuation in the mean. To quantify this deviation, for a given
set of average velocities, we define the deviation $\delta \equiv
v_{max} - v_{min}$, where $v_{max}$ ($v_{min}$) is the largest
(smallest) final average velocity in the set.

\begin{figure}
\includegraphics[width=8.6cm]{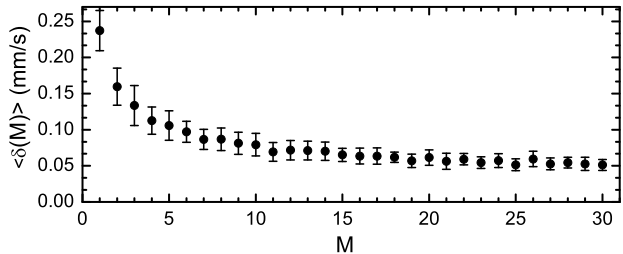}
\caption{(color online) Plot of the average deviation
$<\delta(M)>$ between different ensemble averages as a function of
$M$, the number of runs used to generate the ensemble set. As the
number of independent runs used to generate the ensemble
increases, the deviation between computed averages decreases.}
\label{delta v M}
\end{figure}

Next, in Fig.~\ref{delta v M} we describe the spread in the final
mean values of the velocity profile parameterized by $\delta$ as a
function of an increasing number of independent members of the
time set, $M$, used in constructing the ensemble set. This is done
to test whether or not thirty independent members of the time set
is sufficient to build an ensemble set with physical relevance,
i.e. uncorrelated and independent. One method of testing this is
to consider ensembles constructed from constructed $M < 30$
members of the time set, and measure $\delta (M)$. To increase the
statistics, a number of ensemble sets are generated for each $M <
30$, and one considers $<\delta (M)>$ over these sets.
Figure~\ref{delta v M} is a plot $<\delta(M)>$ for each $M$, and
it confirms our expectation that increasing $M$ should increase
the independence between the randomly selected velocities,
resulting in a better approximation of an ensemble average. For $M
\ge 20$, the value of $<\delta(M)>$ is bounded within range of the
error bar of the value at $M=20$. This has two important
consequences. First, the independence of $<\delta(M)>$ on $M$ for
$M \ge 20$ is strong evidence that thirty independent experimental
realizations provides a sufficient approximation of the ensemble
average for this system. Second, the indication that $<\delta(M)>$
has converged to a nonzero value may be attributed to the finite
size of the system. An interesting future study will be the
dependence of $<\delta(M)>$ as a function of the system size.

We have presented three main results. First, athermal fluctuations
generated during the flow of a bubble raft are sufficient to
produce stationary, time-averaged velocity profiles after a total
strain of approximately one (Fig.~\ref{mean_values}). Second,
these fluctuations do not provide sufficient exploration of phase
space to result in a unique time-average. In other words,
different realizations of a bubble raft produce different
time-averaged velocity profiles that exhibit a measurable and
significant deviation from each other (see Fig.~\ref{mean_values}
and \ref{vel_profiles}). Third, by sampling velocities from
different realizations of the experiment, the resulting ensemble
averages demonstrate a convergence towards similar velocity
profiles (Fig.~\ref{mean_values} and \ref{vel_profiles}). Putting
these results together, one concludes that the limit of time
averaging is different from that of ensemble averaging for this
system in at least one crucial aspect i.e., there is no
convergence to a unique time-average for all experimental
realizations of the bubble raft, but the ensemble averages
converge to final velocity profiles that are consistent with a
unique profile.

\begin{acknowledgments}

This work was supported by a Department of Energy grant
DE-FG02-03ED46071. The authors thank Corey O'Hern and Manu for
useful discussions.

\end{acknowledgments}


\end{document}